\documentclass[prl,aps,epsfig,showpacs,twocolumn]{revtex4}
\usepackage{graphicx}
\usepackage{amssymb}
\usepackage{amsmath}

\begin{document}

\draft

\title{Anomalous training effect of perpendicular exchange bias in Pt/Co/Pt/IrMn multilayers}

\author{Z. Shi and S. M. Zhou}
\affiliation{Applied Surface Physics Laboratory (State Key
Laboratory) and Department of Physics, Fudan University, Shanghai
200433, China}

\date{\today}

\begin{abstract}

\indent A new characteristic is observed in the training effect of
perpendicular exchange bias. For Pt/Co/Pt/IrMn multilayers with
perpendicular magnetic anisotropy, the magnetization reversal
process is accompanied by pinned domain wall motion and the
asymmetry of hysteresis loop is always equal to zero during
subsequent measurements. It is interesting to find that the
exchange field decreases greatly as a function of the cycling
number while the coercivity almost does not change. It is clearly
demonstrated that the training effect of perpendicular exchange
bias strongly depends on the magnetization reversal mechanism of
the ferromagnetic layer.

\end{abstract}

\vspace{5 cm}


\maketitle

\begin{figure}[tb]
\begin{center}
\resizebox*{5 cm}{!}{\includegraphics*{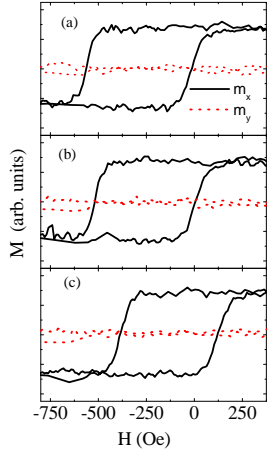}} \caption{Figure
1 Typical out-of-plane hysteresis loops of Pt/Co/Pt/IrMn
multilayers for $n$=1(a), 2(b), and 30(c).} \label{Fig1}
\end{center}
\end{figure}

\begin{figure}[tb]
\begin{center}
\resizebox*{5 cm}{!}{\includegraphics*{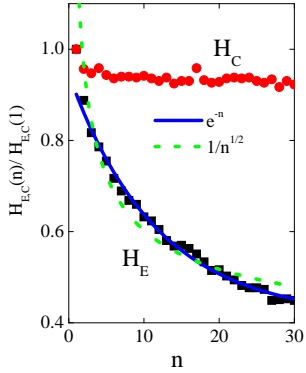}}
\end{center}
\caption{Figure 2 Dependence of $H_{\mathrm{E}}$ (solid square)
and $H_{\mathrm{C}}$ (solid circles) on $n$ under the external
magnetic field along the film normal direction. The solid and dot
lines refer to the fitted results of the exponential function and
the power function, respectively.} \label{Fig2}
\end{figure}

\begin{figure}[tb]
\begin{center}
\resizebox*{5 cm}{!}{\includegraphics*{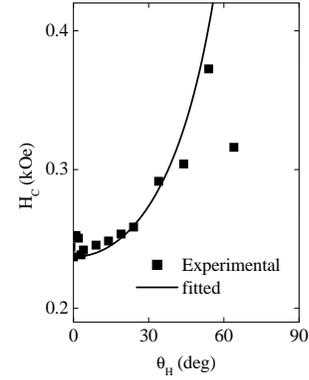}}
\end{center}
\caption{Figure 3 Dependence of measured $H_{\mathrm{C}}$ (solid
square ) on the orientation of the external magnetic field. Here,
the solid line refers to $H_C(\theta_H)\varpropto
H_C(0)/\mathrm{cos}(\theta_H)$.} \label{Fig3}
\end{figure}

\begin{figure}[tb]
\begin{center}
\resizebox*{5 cm}{!}{\includegraphics*{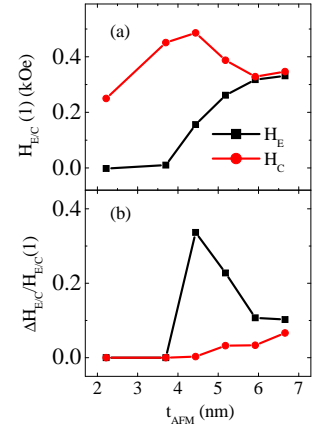}}
\end{center}
\caption{Figure 4 Variations of $H_{\mathrm{E}}$ (solid square)
and $H_{\mathrm{C}}$ (solid circles) (a), $\Delta
H_{\mathrm{E}}/H_{\mathrm{E}}(1)$ (solid square) and $\Delta
H_{\mathrm{C}}/H_{\mathrm{C}}(1)$ (solid circles) with
$t_{\mathrm{AFM}}$ (b)} \label{Fig4}
\end{figure}

\indent As a well known phenomenon, exchange bias (EB) has
received much attention because of its abundant applications in
magneto-electronic devices. In the EB effect, induced by the
magnetic coupling at ferromagnet (FM)-antiferromagnet (AFM)
interfaces, the center of the hysteresis loop is shifted along the
magnetic field axis by an amount of the exchange field
${H_\mathrm{E}}$~\cite{1,2,2a}. Meanwhile, the coercivity
$H_{\mathrm{C}}$ is often increased, in comparison with that of
the corresponding FM layer. Among EB-related phenomena, the
training effect has been studied extensively \cite{3}, in which
${H_\mathrm{E}}$ and $H_{\mathrm{C}}$ are often found to decrease
gradually while cycling the system through successive hysteresis
loops. In most studies on the training effect of either type I or
type II \cite{4}, the coercive field of the descent branch is
shifted greatly while that of the ascent branch does not change
much \cite{5}, and ${H_\mathrm{E}}$ and $H_{\mathrm{C}}$ can
generally be fitted by a linear function of $1/\sqrt{n}$ \cite{1}.
Moreover, most of studies have focused on the training effect of
the in-plane EB. In contrast, the training effect of the
perpendicular EB has been less well understood \cite{6}. To
elucidate the EB training effect, various theoretical models have
been proposed \cite{7,7a}. In these models, effects of temperature
and AFM layer thickness $t_{\mathrm{AFM}}$ on the EB training
effect are considered. Up to date, however, the effect of the
magnetization reversal mechanism on the training effect has been
neglected. In this Letter, we have studied the training effect of
the perpendicular EB in Pt/Co/Pt/IrMn multilayers, in which the
magnetization reversal process of either branch is accompanied by
domain wall motion. Although ${H_\mathrm{E}}$ shows a prominent
training effect, $H_{\mathrm{C}}$ almost does not change with
cycling number $n$. This new feature clearly indicates that the EB
training is strongly
related to the magnetization reversal mechanism.\\
\indent  A large specimen of Pt(3 nm)/Co(0.5 nm)/Pt(0.5 nm)/
Ir$_{\mathrm{25}}$Mn$_{\mathrm{75}}$(=IrMn) was deposited on thin
glass substrate by magnetron sputtering at ambient temperature,
where the IrMn layer thickness takes a wedge shape to avoid the
run-to-run error. Uniform multilayer of Pt(3 nm)/Co(0.5 nm)/Pt(0.5
nm)/IrMn (4 nm) was also grown. The base pressure of the system is
$3.2\times10^{-5}$ Pa and the Ar pressure 0.40 Pa during
deposition. A magnetic field of about 250 Oe was applied normal to
the film plane in order to establish the perpendicular EB.
$m_{\mathrm{x}}$ and $m_{\mathrm{y}}$ signals were recorded
simultaneously by vector vibrating sample magnetometer (VVSM) of
model 7407 from LakeShore Company. All measurements were
performed at room temperature.\\
\indent Figure~\ref{Fig1} demonstrates typical out-of-plane
hysteresis loops of a uniform sample Pt(3 nm)/Co(0.5 nm)/Pt(0.5
nm)]/ IrMn(4 nm) with \textit{n}=1, 2, and 30. Apparently, the
perpendicular EB is well established. ${H_\mathrm{E}}$ and $H_C$
are 289/265, 257/254, and 131/247 in the unit of osterds for
$n$=1, 2, and 30, respectively. Two distinguished characteristics
can be found. At first, ${H_\mathrm{E}}$ decreases with $n$ but
$H_{\mathrm{C}}$ does not change much. During subsequent
measurements, both the coercive field of the descent branch and
that of the ascent branch move towards the positive magnetic
fields~\cite{4,5}. Remarkably, the amount of the field shift for
the ascent branch is almost equal to that of the descent branch,
unlike convention results in which the amount of the field shift
for the ascent branch is much smaller than that of the descent
branch. Secondly, for the present Pt/Co/Pt/IrMn multilayers, the
asymmetry of hysteresis loop is equal to zero and does not change
during subsequent measurements, as revealed by the curve of
$m_y\,$-$H$. It is quite different from the conventional results
in which the asymmetry changes greatly after the first
magnetization reversal~\cite{7}.\\
\indent Figure~\ref{Fig2} shows the dependence of ${H_\mathrm{E}}$
and $H_{\mathrm{C}}$ on $n$. ${H_\mathrm{E}}$ decreases gradually
with increasing $n$. $H_{\mathrm{E}}(n=30)$ is reduced to be less
than half of $H_{\mathrm{E}}(1)$. For comparison, it is fitted by
a linear function of both $1/\sqrt{n}$ and $e^{-\mathrm{n}}$. One
can find that the exponential function can better fit the measured
results, except for $n$=1. More remarkably, $H_{\mathrm{C}}$ does
not change for $n>2$, except for the first drop, which has been
explained as a result of the spin flop in the AFM layer~\cite{7}.\\
 \indent At first, it should be pointed out that the new feature
 of the training effect is not intrinsic properties of the perpendicular
 EB. This is because in perpendicularly exchange-biased Co/Pt/CoO
 multilayers, $H_{\mathrm{C}}$ is also found to shrink during subsequent
 measurements of hysteresis loops \cite{6}. As analyzed below, above salient
 phenomena are induced by the unique magnetization reversal mechanism.
 As shown in Fig.~\ref{Fig1}, the magnetization reversal process for either the
 descent branch or the ascent one is accompanied by the domain wall
 motion because there is no magnetic component perpendicular to the
 external magnetic field. This can also be convinced by the curve of
$H_{\mathrm{C}}$ versus the orientation of the external magnetic
field $\theta_{\mathrm{H}}$ as shown in
 Figure~\ref{Fig3}. $H_{\mathrm{C}}$ initially increases as the external magnetic field is deviated
 from the normal direction and then decreases sharply, which can be fitted by the
 modified Kondorsky model~\cite{13}. A maximal $H_{\mathrm{C}}$ is
 located at the critical angle $\theta_{\mathrm{HC}}$ of 60 degrees, which is
 determined by the competition between the pinning field of domain wall and
 the uniaxial anisotropic field. Similar results were
 obtained in other films \cite{10,11,12} and attributed to the
 hindered domain wall motion. As the pinning field is smaller than the
 uniaxial anisotropic field at $\theta_H$ smaller than $\theta_{\mathrm{HC}}$, the magnetization
 reversal process is accompanied by the pinned domain wall motion and $H_C(\theta_H)\varpropto H_C(0)/\mathrm{cos}(\theta_H)$,
 in which $H_{\mathrm{C}}(0)$ refers to the coercivity at the direction normal to the film plane.
 As the pinning field is larger than the
 uniaxial anisotropic field at $\theta_H$ larger than
 $\theta_{\mathrm{HC}}$, the coherent magnetization rotation occurs and the coercivity decreases with
 increasing $\theta_{\mathrm{H}}$. In experiments, we found that for Co/Pt
 multilayers, the magnetization reversal process is also accompanied by the
 domain wall motion. Therefore, Pt/Co/Pt and Pt/Co/Pt/IrMn mutilayers have the same
 magnetization reversal mechanism.  \\
 \indent With exchange coupling at the FM/AFM interface, the AFM spins are dragged by
 the exchange field from the FM layer. With pinned domain wall motion, the FM
 magnetization and thus the exchange field exerted on the AFM
 spins are either parallel or antiparallel to the pinning direction during
 measurements of hysteresis loop. Accordingly,
 the AFM spins can only be switched by 180 degrees \cite{14} and the number of the
 AFM spins parallel to the pinning direction is changed, leading to the training effect of ${H_\mathrm{E}}$.
 For the present FM/AFM multilayers, the coercive field at
 $\theta_{\mathrm{H}}=0$ for the ascent (+) and the
 descent (-) branches are equal to the effective pinning fields, i.e,
 $H_{\mathrm{C}}(\pm)=-H_{\mathrm{E}} \pm{1 \over M_{\mathrm{FM}}}
(\partial \gamma /\partial x)_{\mathrm{max}}$,
 where the domain wall energy $\gamma$ consists of two parts, including the
 domain wall energy of the FM layer and the interface energy. The former
 one corresponds to the intrinsic coercivity of the FM layer and the latter one to
 the coercivity enhancement. Apparently, the inhomogeneity of the latter
 one does not change during the sweeping of the magnetic field. It is
 different from the case of the magnetization coherent rotation during the
 magnetization reversal process, in which the coercivity enhancement is
 contributed from the effective uniaxial anisotropy
 induced by rotatable AFM grains\cite{15}. During the sweeping of the magnetic field, the
 switching of AFM grains is triggered from the non-equilibrium orientation
 to equilibrium one. Less AFM grains can be rotated after subsequent
 measurements and the coercivity is reduced.\\
\indent Figure~\ref{Fig4} shows the dependence of
$H_{\mathrm{E/C}}\ (n=1)$ and the training effect on
$t_{\mathrm{AFM}}$. At $t_{\mathrm{AFM}}$ smaller than 3.6 nm,
${H_\mathrm{E}}(n=1)$ is equal to zero. With increasing
$t_{\mathrm{AFM}}$, it increases sharply and finally approaches
saturation. Meanwhile, $H_{\mathrm{C}}$ acquires a maximum at
$t_{\mathrm{AFM}}$=4.5 nm. As $t_{\mathrm{AFM}}$ is increased,
$\Delta H_E/H_E(1)$ inclines to the maximum sharply and then
decreases dramatically whereas $\Delta H_C/H_C(1)$ is negligible
for all $t_{\mathrm{AFM}}$. These results can be understood using
a simplified thermal activation model \cite{15}. At small
$t_{AFM}$, the energy barrier (the product of the volume and the
anisotropy energy) is too small to overcome the thermal energy and
thus all AFM grains are in the superparamagnetic state, resulting
in zero ${H_\mathrm{E}}$ and small $H_{\mathrm{C}}$ and negligible
training effect. At the intermediate $t_{\mathrm{AFM}}$, some AFM
grains become thermally stable, which results in non-zero
${H_\mathrm{E}}$, the $H_{\mathrm{C}}$ enhancement, and the
prominent training effect of ${H_\mathrm{E}}$. At large
$t_{\mathrm{AFM}}$, the energy barrier is high enough to overcome
the thermal energy, ${H_\mathrm{E}}$ is saturated and
$H_{\mathrm{C}}$ is decreased.
At the same time, the training effect of ${H_\mathrm{E}}$ is suppressed.   \\
\indent In summary, the Pt/Co/Pt/IrMn multilayers were fabricated
and the perpendicular EB is established. For both single FM and
FM/AFM multilayers, the magnetization reversal is accompanied by
the domain wall motion. The asymmetry of hysteresis loop is always
equal to zero during subsequent measurements. The exchange field
decreases greatly as an exponential function of $n$ while the
coercivity does not change. It is clearly demonstrated that the EB
training effect strongly depends on the magnetization reversal
mechanism of the FM layer. \\
 \indent Acknowledgement This work was supported
by the National Science Foundation of China Grant Nos. 50625102,
10574026, and 60490290, the National Basic Research Program of
China (2007CB925104) and 973-Project under Grant No. 2006CB921300,
Shanghai Science and Technology Committee Grant No. 06DJ14007.

\end{document}